\documentclass{elsart}

\usepackage{natbib}

\usepackage{graphicx}

\usepackage{amssymb}

\usepackage{lineno}

\begin{document}
 
\begin{frontmatter}
\journal{Planetary \& Space Science}
\volume{57}
\firstpage{1218}
\lastpage{1227}


\title{That's the Way the Comet Crumbles: Splitting Jupiter-Family Comets}


\author{Yanga R. Fern\'andez\corauthref{cor}}
\corauth[cor]{Corresponding author. Tel.: +1 407 8236939;
fax: +1 407 8235112.}
\ead{yan@physics.ucf.edu}

\address{University of Central Florida, Department of Physics, 4000 Central
Florida Blvd., Orlando, FL 32816-2385 U.S.A.}

\begin{abstract}

Our current understanding of split, Jupiter-family comets is reviewed.
The focus is on what recent studies of comets have told us about
the nature of the splitting phenomenon. The goal is to not repeat
the information given in recent reviews of split comets, but to
build upon it.  In particular, we discuss comets that have suffered
splitting or fragmentation events in the past few years. These
include comets (a) 57P/du Toit-Neujmin-Delporte, observed with a
long train of fragments in 2002; (b) 73P/Schwassmann-Wachmann 3,
which split in 1995 and was extensively studied during its relatively
close passage to Earth in 2006, during which dozens of fragments
were discovered and studied;  and (c) 174P/Echeclus, a Centaur and
potentially future JFC, which split in late 2005 and was the first
such Centaur observed to do so.  We also discuss recent observations
by SOHO of split comets that are likely of short-period.  
The Spitzer Space Telescope has
observed many JFCs and provided us with unprecedented detailed views
of cometary debris trails, which may be thought of as a middle
ground between ``normal"  ejection of micron-sized dust grains and
the cleaving off of meter-to-kilometer sized fragments. We will
also discuss potential breakthroughs in studying splitting JFCs
that may come from future surveys.

\end{abstract}

\begin{keyword}

comets -- splitting \sep comets -- individual 
(57P, 73P, 174P) \sep comets -- evolution 


\PACS 96.30.Cw \sep 96.30.C-

\end{keyword}

\end{frontmatter}

\newpage
\firstpage{2}

\section{Introduction}

The ``typical" Jupiter-family comet (JFC) loses mass through a
relatively slow process of volatile sublimation and dust entrainment.
Common mass loss rates when a JFC is near perihelion are roughly
$\sim10^{1 - 3}$ kg/s \citep{ahe95}, depending on the size of the
nucleus's ``active area." This indicates that nominally the cometary
nucleus could disintegrate away probably only after thousands of
orbits around the Sun. This end-state of a JFC has never been
directly observed, although such very tiny comets (meters in diameter)
that are about to disintegrate would be difficult to discover. In
any case current thinking \citep[e.g.][]{ms04} holds that most JFCs
end their lives either by plunging into the Sun, colliding with a
planet, simply turning off and becoming asteroid-like, or  by
catastrophic fragmentation.

This review will focus on this last option and its attendant large
mass loss rate. The primary motivation for studying split comets
is that they are laboratories for understanding cometary structure
and bulk mechanical properties.  Since this can give us clues about
how comets are put together, split comets can be an important window
for investigating the details of planetary formation and specifically
the accretion of solids into icy planetesimals. As mentioned 
above, split comets also represent an aspect of cometary
evolution. Since split comets expose previously buried material to
the space environment and thus to our telescopes, they give us a
way to probe chemical and thermophysical changes in cometary nuclei.

In this paper, recent developments in our understanding of split
comets will be discussed.  An excellent recent review of this topic
is provided by \citet{boehnh04}, and there have been several likewise
good reviews before that \citep[e.g.][]{seka82,hm92,sek97,boehnh02}.
The goal of this paper is to not duplicate that earlier work but
rather to provide updates on work occuring in the interim.

\section{Observational Aspects of Splitting}

Conceptually, the question of ``how do you know when a comet has
split?" is easy to answer.  The observational manifestation of a
splitting is a condensation that appears away from the head of the
comet but moving with very  nearly  the same proper motion.  However
not all near-nuclear condensations are actually indicative of a
split; the condensation may or may not hold a solid body and may
be simply (e.g.) a trick of perspective on  jet features in the
coma, or a clump of dust from an outburst.  The overall dust
production rate and the complexity of  coma morphology can make the
actual identification of a real splitting  problematic.  In particular,
a splitting is often associated with an outburst or a brightening,
potentially making it even more difficult to identify a fragment.

Interestingly, a fragment may not appear even when a  comet brightens
dramatically.  For example comet 29P/Schwassmann-Wachmann 1
has frequent outbursts \citep[see, e.g.,][]{jewitt90}, 
yet no fragment has ever been seen.

On the other hand a comet that appears perfectly ``typical" could
have had a recent splitting,  but the fragments may be more than a
few arcminutes from the head, i.e. beyond the field-of-view of a
typical CCD camera.  Due to sporadic monitoring of most comets, a
fragment will not always be seen in the near-nuclear region; it may
not be discovered until well after the split, as was the case for
fragment B of 57P/du Toit-Neujmin-Delporte \citep{mars02a}, which
was  found 0.2$^\circ$ away from fragment A, the comet's main head.
The even more extreme case is the paternity of comets 42P/Neujmin
3 (discovered in 1929) and 53P/van Biesbroeck (discovered in 1954),
which were found \citep{car85}  to be pieces of one comet that split
after a close approach to Jupiter in 1850.

Another problem in determining whether a split has happened is
simply faintness. A 50-meter radius bare nucleus that is 2.0 AU
from the Sun and 1.0 AU from Earth -- all reasonable numbers for a
cometary fragment -- will have an R-band magnitude of about 25
(for a geometric albedo of 0.04),
which is beyond the reach of many facilities. Even when such a
fragment is actively outgassing and thus brighter, the observation
itself may not be done in such a way as to detect the fragment.

All this makes determining the frequency of fragmentation among
JFCs relatively difficult to measure. A list of published
instances of JFC fragmentation is give in Table 1, but since many
comets have long intervals (months or years) where  no observations
are obtained by either professionals or amateurs, this is naturally
a lower limit of the true roster.  \citet{hm92} estimated that a
JFC has 0.3\% chance of splitting per perihelion passage, based on
the historical record of observed fragmentations. \citet{cj94}
observed a sample of 34 JFCs with CCDs in the late 1980s and early
1990s, and found 2 had split, corresponding to a $\sim$1\% chance
of a JFC splitting per year.  If these are accurate estimates of
the splitting rate, one remarkable consequence  is that over the
course of the $\sim10^3$ orbits that a typical JFC will be active
\citep{ld97}, it can expect to split perhaps dozens to hundreds of
times. As noted by \citet{cj94}, while JFC splittings  may be
perceived as being rarer than splittings by near-isotropic comets,
a given JFC will shed fragments many times during its active life.

This finding suggests that the shape and rotation state of a JFC
is intimately tied to the specific fragmentation events it has
suffered, since the amount of mass coming off in a fragment can be
significant compared to the total mass lost by normal outgassing
in the course of an orbit.  Suppose a JFC with perihelion at 1 AU and
aphelion at 5.2 AU (Jupiter's distance) has a mass loss that
is 200 kg/s at perihelion and is proportional to the 
inverse-square of
heliocentric distance. This comet will lose about $4\times10^9$ kg
in one orbit.  If the effective radius of the nucleus is 2 km, and
the density is about 400 kg/m$^3$, that is just 0.03\% of the comet's
total mass. If the comet is ``active" over 10\% of its surface, the
mass loss from those active areas erodes about 2 m deep. If an
equivalent mass were to come off as a spherical fragment, however,
the fragment would be 130 m in radius. This is a reasonable size
for a ``typical" fragment of a JFC \citep{boehnh04}.  Such a splitting
would represent a significant change in gravitational field, angular
momentum, and topography.

As mentioned earlier, we have not discovered a population of very
tiny comets.  We do not have observational evidence of a population
of deka- or hectometer sized JFCs that could be made up of fragments.
The discovery biases almost certainly play a role in this, although,
as \citet{mhm04} showed, if one accounts for the biases it seems
that the JFC size distribution really does fall off below $\sim$1
km. In other words if the JFC size distribution were a power-law,
we should have discovered more comets smaller than 1 km than we
actually have. This suggests that most deka- or hectometer sized
JFC fragments will not stay as  coherent bodies for long
(years to decades) but eventually disintegrate.

In any case, directed, deep searches for smaller JFCs would be
useful so that we could directly address this problem with as little
observational bias as possible. In particular, a search that covers
the projected orbits of several JFCs would put constraints on the
lifetimes of the fragments.  For example, in Fig. 1 we show a Spitzer
Space Telescope mosaic of the vicinity of comet 73P/Schwassmann-Wachmann
3 in May 2006. Approximately three dozen fragments can be seen; how
many will survive to the comet's next perihelion in 2011?  (We
discuss this comet further in \S 4.2.)

\section{Causes of Splitting}

While tidal disruption as a mechanism for splitting comets is well
understood, it is responsible for only a small fraction of observed
splits. Briefly, a comet passing close enough to a planet (usually
Jupiter) or the Sun will feel different gravitational forces on one
end of its nucleus compared to the other. The difference in forces
can be strong enough to overcome the body's cohesion.
The extremely
low tensile strength and high porosity of cometary nuclei, as
suggested most recently by the Deep Impact visit to comet 9P/Tempel
1 \citep{es07,hh07,ah08}, indicate that a tidal force need not be
that strong to successfully rip a comet apart. However, in only one
case -- comet D/1993 F2 (Shoemaker-Levy 9)-- 
are we very sure that tidal disruption
is the cause of the fragmentation.  Comet 16P/Brooks 2 probably also
suffered this fate but it was not discovered until a few years after
the purported close-approach.

For all the other comets in Table 1, tidal disruption could not
have been the cause of the split. In these cases, we are no closer
to understanding why a particular comet splits when it does than
we were when \citet{boehnh04} wrote his review. He gives four other
methods that could cause splitting: by fast rotation, by thermal
stress, by internal gas pressure, and by impacts. Since that review
was written, our understanding of the thermal properties of cometary
nuclei has improved as a result of the Deep Impact experiment. For
example there is stronger evidence now that the thermal conductivity
of cometary nuclei is extremely low \citep{sun07,grou07}, so low
that the thermal pulse can penetrate only a few centimeters due to
diurnal heating. This means that on diurnal time scales 
thermal energy may not be transported effectively
into the comet's interior any faster  than the surface can itself
be excavated  by normal  cometary activity.  The existence
of abundant CO$_2$ -- with its low sublimation temperature -- in
P/Tempel 1's coma \citep{fea07} also indicates that the comet could
not have been entirely baked out.  Of course these are results for
just one comet, but if they are indicative of the ``average" thermal
properties of a JFC, then perhaps it may be less likely for JFCs
to split as a result of thermal stress. On the other hand, the low
strength of a JFC means that one can conceive of localized structures
(e.g.  sheer cliffs in depressions or concentrations of less porous
rock)  where  thermal stress in just a small volume could cause a
much bigger volume to  break off.  This relates to more fundamental
questions of nucleus structure about which we do not yet have much
data.  How well are the volatiles and refractories mixed within the
nucleus? Is the high porosity manifest in micro- or macro- scales?
How common is significant topography?

\section{Recent Results}

\subsection{Comet 57P/du Toit-Neujmin-Delporte}

Fragment B was discovered in July 2002 well away from the main part
of the comet. Soon after, we \citep{fern02} discovered eighteen
more fragments (named ``C" through ``T") along the line of variation
(the projected orbital path) and extending out 27 arcminutes from
the comet's head. A montage of these individual fragments as seen
on UT July 17 is shown in Fig. 2, along with a plot of
the fragments' positions.  The fragments had varying
brightnesses ranging from 20 to 23.5 mag in R-band. They also had
widely varying degrees of condensation;  some fragments were nothing
more than blobs of dust with no central source, such as I and P.
All fragments seemed to be actively outgassing, though with apparently
varying production rates.

Interestingly, there was no outburst in 2002 associated with this
shedding of mass. There was a significant outburst at the previous
apparition, in 1996, when the comet was observed to be about 5 mag
brighter than expected. However dynamical analysis of the largest
fragments by \citet{sc02} suggests that they could not have broken
off six years earlier.

The fragments must represent a significant fraction of the comet's
total mass. How much mass, and what fraction is it?  The size of
the nucleus before fragmentation is unknown, but \citet{lf01} derived
an upper limit to the radius of 1.1 km.  So for an assumed density
of 400 kg/m$^3$, the upper limit to total mass would be approximately
$\sim2\times10^{12}$ kg.  The activity of the fragments makes their
size estimation problematic, so a mass estimate is even more
difficult. However we can make an order-of-magnitude analysis.  For
example, fragment G had an R-band magnitude of about 22.5, which
at the distance of 57P at the time, would correspond to a solid
body with effective radius of about 130 meters.  Such a body alone
would be $\sim$0.1\% of 57P's total mass.  This is an overestimate
for this particular fragment since cometary dust contributes to the
magnitude, but there are a few fragments that are likely much bigger
than G, so this is a reasonable order-of-magnitude estimate of 57P's
mass loss.

A slightly more sophisticated analysis would be to na\"ively convert
all 19 fragment magnitudes to sizes, and thus create a cumulative
size distribution (CSD).  One finds that such a distribution is
fairly shallow, with $N(>\!R)$, the number of objects with radius
bigger than $R$, proportional to only $R^{-1.3}$. While this totally
ignores the obvious activity of the fragments, if activity scales
roughly with the fragment's surface area, the power-law slope of
the size distribution will be unaffected. This slope means that the
total mass $m$ contained in fragments up to some size cutoff $R_0$
is proportional to $R_0^{1.7}$.  This suggests that much of the
fragmented mass is in the largest pieces such as B, E, and F.

Interestingly, the CSD slope is shallower than that of large dust
grains found in JFC trails, as measured by \citet{reach07a}. They
found $N(>\!R) \propto R^{-2.3}$ for grains over 0.25 mm.  Whether
this is a clue as to the physical mechanism behind fragmentation
remains to be studied.

The real significance of the discovery of the train behind 57P is
that it was the first time so many fragments had been observed
around a \underbar{surviving}, \underbar{non-tidally} \underbar{disrupted}
comet. Other comets have had as many fragments  but only while being
completely broken apart or after passage by Jupiter or the Sun
(e.g., comet C/1999 S4 (LINEAR), comet D/1993 F2 (Shoemaker-Levy 9),
and the Kreutz sungrazers).   It motivates
the question of whether shedding events happen more frequently than
thought and are just being missed due to the extreme faintness of
the fragments. Indeed, the sheer length of 57P's train suggests
that there were several fragmentation episodes in the past.  If a
JFC loses only fragments of magnitude 23, 24, or fainter, only deep
imaging of the comet during the course of monitoring will reveal
them.

The case of 57P also raises questions about the endurance of the
smallest fragments.  \citet{sc02} state that fragment F  would have
left the primary nucleus -- i.e. fragment A -- 
about 14 months before the discovery observation; at
discovery, F was 6 arcmin from Fragment A.  They predicted the
future motion of fragment F based on this model, but unfortunately
apparently little if any data could be obtained by observers to
corroborate or refute the hypothesis. However, if true, it suggests
that large fragments farther down the train could have been released
at even earlier times. Fragment T is 27 arcminutes from fragment A
-- 4.5 times farther than fragment F. T could have left the main
nucleus (on its own or as part of another fragment) years in the
past.  This would not necessarily be unprecedented, since as
\citet{boehnh04} notes JFC fragments can have long endurance (hundreds
of days), and there is indication that JFCs can split most anywhere
along their orbit.  The estimation of the endurance of fragment T
would also depend on at what relative speed it left the nucleus.
In any case, fragment T is a fairly small fragment: could it have
survived for years, and  are we just now seeing the end of its life?
Is it really persistent despite its small size? It is also possible
that fragment T is a subfragment of a brighter fragment, say fragment
S. Even so, the distance between S and T is appreciable (3.2 arcmin)
and suggests an endurance for T of several months despite its small
apparent size.  \citet{bn01} have modeled the survival times of
fragments and show that small, clean icy fragments can survive for
months after separation, although whether or not fragment T can be
clean ``enough," is uncertain. The images in Fig. 2 show cometary
dust, not gas, and so there must be some ``dirtiness" to the
fragments, which would shorten their lives \citep{hann81,lien90}.

The faintness of the fragments, and then later the fact that the
comet was post perihelion and at unfavorable elongation, prevented
detailed follow up.  The endurance of the the fragments was not
directly measured. Therefore further observations of the fragments
and of the main comet itself at future apparitions would be extremely
useful to understand how this comet is evolving.

\subsection{Comet 73P/Schwassmann-Wachmann 3}

One of the most-widely observed split comets since D/Shoemaker-Levy
9 has been 73P/Schwassmann-Wachmann 3, which first
split in 1995 and approached to within 0.07
AU of Earth in May 2006.  This recent visit was a monumentally 
important apparition
scientifically since it would allow us detailed studies of bright
and relatively-fresh fragments -- fragments whose ices had formerly
been deeply embedded within the comet's nucleus but were now exposed
to sunlight. In a way, it was similar to the Deep Impact experiment
in that 73P provided us with a close-up view of subsurface pristine
cometary material.

A review of this comet's behavior was given by \citet{betal02} and
by \citet{sek07}, which we briefly summarize here.  This comet split
into several fragments in 1995, some of which deactivated or
disintegrated before the apparition was over. At the comet's next
apparition in 2000 and 2001, two of the 1995 fragments, B and C
(the primary part of the comet), were recovered, and two new
fragments, E and F, were confirmed (although apparently E broke off
from C during the previous orbit). F disappeared later in the
apparition but the stage was thus set for the 2006 apparition: would
B and C survive another perihelion passage? Would E have survived?
Would there be a train of even more fragments?

The result was that in April and May 2006 over 60 fragments were
found  -- including B and C but not E.  The proximity to Earth
certainly helped; a fragment of the same size as 57P-G (mentioned
in the last subsection) would be about 3 mag brighter at
73P's geocentric distance.  
In any case some of the fragments were very short-lived,
lasting for only a few days. Amateur astronomers contributed greatly
to the census of fragments, in fact even finding fragments that
disappeared within a day and so could not be followed up and formally
named. Lists of the fragments and relevant info have been compiled
by \citet{sek07} and \citet{birt08}, and the former also discusses
some preliminary work on describing the cascading fragmentation.

One of the many spectacular images from this apparition is shown
in Fig. 3, which comes from the Hubble Space Telescope.  Fragment
B was a dynamic and rapidly changing fragment for much of the
apparition, and this image epitomizes this. It shows several
subfragments (a.k.a. ``mini-comets") tailward of fragment B itself.
These pieces are probably dekameter in scale, and a sequence of
images from HST show these fragments moving down the tail, outgassing
until they disintegrate away in timescales of only hours or days.
The image field-of-view is only about 25 arcseconds. A wider scale
picture is shown in Fig. 1, covering almost 5 degrees of sky along
the comet's orbital path.  The dynamics of the small fragments in
Figs. 1 and 3 have been analyzed by \citet{reach07b}, who found
that the HST fragments are strongly affected by the non-gravitational
reaction force due to outgassing -- suggesting a high volatile
content. On the other hand, the ``meteoroids" in the Spitzer image
are moving as would be expected simply from radiation pressure and
solar gravity -- suggesting a low volatile content. This perhaps
could be explained as an evolutionary effect, where the HST fragments
dry out to become the Spitzer fragments. This does require that the
HST fragments would have to have sufficient size and sufficiently
low rock-to-ice ratio to be able to survive for the several days that 
they are seen in the HST data. Unfortunately, identifying
the fragments seen by HST on April 18 with some part of
the extended emission seen
in the Spitzer mosaic on May 4 would be problematic since the spatial
resolutions are so different and the time gap is so large. 

The variability of fragment B is in stark contrast to fragment C,
the primary fragment \citep{sek07}.  That fragment remained relatively
stable, with only slowly varying activity, nowhere near the frequency
and amplitude of changes seen from night-to-night (and sometimes
within a night) in fragment B. The two fragments must be roughly
comparable in size, yet the specifics of B's shape and the location
of its volatiles has made apparently a huge difference in the
evolution.

Many observations of comet 73P were obtained during its apparition,
and analysis on these rich datasets continue. We summarize here
some of the exciting results and apologize for oversights.

Arguably some of the most important findings involve the composition.
Overall, 73P seems to be depleted in CH$_3$OH, C$_2$H$_6$, and
C$_2$H$_2$, but has ``typical" abundance of HCN \citep{vill06,dis07,kob07}.
This means that this ecliptic comet that came from the trans-neptunian
region is compositionally similar to C/1999 S4 (LINEAR), a carbon-chain
depleted Oort Cloud comet \citep{dm08}. Furthermore it is unlike
its fellow ecliptic comet 9P/Tempel 1 \citep{dm08}. This matching
of compositions across dynamical classes hints that there was
sufficient mixing in the protoplanetary disk to allow individual
icy planetesimals to accrete material from various regions. In other
words perhaps there is not necessarily a compositional distinction
that exactly matches the dynamical distinction of ecliptic comets
forming beyond Neptune and Oort Cloud comets forming among the giant
planets. Alternately, there could simply be interlopers polluting
the  dynamical groups. In particular, we are only now starting to
build up a statistically significant sample of the parent-molecule
composition of JFCs \citep{mumma08}
to build upon the daughter-species work by
(e.g.) \citet{ahe95} and \citet{sb08}.  
We can note that 73P is a ``depleted" comet
in the \citet{ahe95} taxonomy and 9P is ``typical;" as the parent
species of more JFCs are observed, 73P and 9P can be placed into
better context.

Several people compared the two main fragments, B and C, to each
other \citep[e.g.][]{biv06,vill06,schl06,ndr07,kob07}.  The consensus
result is that B and C have similar composition.  This is an important
finding since both fragments are relatively large fractions of the
original comet and so show us fresh material that formerly was very
deep inside the nucleus.  They should be excellent laboratories for
determining heterogeneity -- i.e., whether large blocks of the comet
have different compositions. No such effect was found, and this is
in stark contrast to comet 9P, where the heterogeneity was quite
obvious after the Deep Impact flyby \citep[e.g.][]{fea07}. The fact
that 73P seems to be both (a) homogeneous and (b) different from
what may be currently considered ``typical" composition suggests
that this comet had an atypical formation history. However this is
speculative and, again, we are suffering somewhat from a
small JFC sample. In any case we clearly see the vital need for
more surveys of JFC composition.

One important additional result regarding composition was presented
by \citet{schl06}. They showed that the B and C fragments both have
the same ``depleted" abundance of CN, C$_2$, and C$_3$ relative to
water. As mentioned, since the fragments were outgassing relatively
pristine material, this suggests that this depletion seen in many
JFCs could not be an evolutionary effect, but rather is primordial.
It may be indicative of the chemistry happening in the protoplanetary
disk  at the location where at least some of the JFCs formed.

The proximity of 73P motivated many observers to investigate the
physical properties of the fragments.  Of primary concern were the
fragments' sizes and masses.  \citet{boehnh99} observed the comet
in 1994, before breakup, and estimated an upper limit of 1.1 km for
the radius.  Some of the most exciting data on the nuclei after
break-up were obtained by \citet{how07}, who used radar to obtain
Delay-Doppler maps of fragments B and C in May 2006.  73P was at
the time only the second comet so imaged. The data show that fragment
B is at least 0.2 km in radius and that fragment C is about 0.5 to
1 km in radius. These results are consistent with earlier estimates
of those two fragments' sizes \citep{boehnh02,toth03,toth05,toth06}.
While fragment C is the primary remnant of the comet, fragment B
took a significant fraction of the mass with it.

Knowing the rotational states of the larger fragments could potentially
give insight into the fragmentation process, the dekameter-scale
structure of the comet, and/or the cometary mass. So far, there
have been reports only for fragment C \citep[e.g.][]{storm06,toth06}.
Fragment B was so active and changing on such short timescales that
obtaining either a photometrically- or morphologically-derived
period may be challenging. Coma structures were seen however
\citep[e.g.][]{bonev08} so a sufficient baseline of observations
could prove fruitful.  Rotation periods for the other relatively
bright fragments, such as G and H, have not been reported to our
knowledge.

Images such as Figs. 1 and 3 make it clear that there is a continuum
of sizes among the fragments. An analysis of the size distribution
of all fragments has yet to be presented, but \citet{fuse07} have
studied a group of fragments that at the time had all just recently
broken off fragment B.  Their processed image is
shown in Fig. 4, and was obtained a few weeks after the HST image
in Fig. 3. They identify 54 fragments in their data, all of which
were active, and measured the luminosity of each.  Assuming that
the activity is proportional to the fragment surface area, they
then derive a CSD power-law slope\footnote{Note that \citet{fuse07}
call their ``$q$" the CSD power-law slope but it is actually the
differential size distribution's slope; their $1-q=-1.1$ is the CSD
slope.} based on their 54 fragments: $N(>\!R) \propto R^{-1.1}$.
This is tantalizingly similar to the rough CSD slope for 57P as
discussed in \S 4.1.  Both CSD slopes are much shallower than the
overall JFC CSD slope as derived by \citet{mhm04}, suggesting that
the process of splitting has an underlying physical mechanism that
is different from the primary evolutionary processes (such as
collisions and erosion) that affect kilometer-scale JFCs. 
It is important to note however that there is a size mismatch, so such a
conclusion is tentative; there are no independent JFCs known to have
nuclei as small as the fragments seen in 73P and 57P. Interestingly,
both \citet{mhm04} and \citet{sama07} have used simulations
to suggest that the JFC
CSD becomes shallower at sub-kilometer sizes due to a real
dearth of such comets in the inner Solar System.

\subsection{Comet 174P/Echeclus}

The Centaur 174P/Echeclus = (60558) 2000 EC$_{98}$ was discovered
by Spacewatch in March 2000 \citep{sco00}, and it orbits between
5.9 and 15.6 AU from the Sun. Its current orbital intersection
distance with Jupiter is 0.9 AU and with Saturn is just 0.2 AU; as
a Centaur it is likely to be significantly perturbed on $\sim$10$^7$
year timescales \citep{ld97} and may become a JFC.  No cometary
activity was reported for several years after its discovery, and
physical properties of the bare object were obtained by several
groups  \citep{rou05,lr07,stansb08}.  Activity was first noticed
in December 2005 by \citet{choi06a} and continued through May 2006
\citep{choi06b,weissman06}, while the comet was about 13 AU from
the Sun.

\citet{bauer08} present an analysis of contemporaneous ground-based
visible and Spitzer infrared imaging of the comet from February
2006. Their images are shown in Fig. 5, and demonstrate why this
comet should be counted as one that has split. The center of
brightness of the cometary activity is not on the nucleus itself,
but offset (by six arcseconds at the time the images were taken).
Furthermore there is a  condensation embedded in the coma.  However,
imaging obtained in March 2006 (i.e. the following month) by
\citet{rou08} shows a more diffuse coma, and they state that the
surface brightness profiles of the coma suggest that the dust is
no longer coming from a central source but rather from a diffuse
source.

One hypothesis to explain the observations is: Echeclus itself was
mostly inactive, but perhaps active enough in one locale for a
fragment to break off. This fragment stayed active, while the
remainder of Echeclus continued to have no activity. After a few
months, the fragment itself began to disintegrate into smaller
pieces or subfragments.  The subfragments remained active and so
-- at the spatial resolution obtainable from Earth -- the coma
appeared to emanate from a distributed source.  The problem with
this scenario is explaining why there would not be activity from
the ``hole" on the primary created by the departing fragment.

An alternate hypothesis is that we are seeing a satellite of Echeclus
that just happens to be active. However, the motion of the fragment
over the course of several months suggests that it is moving
hyperbolically \citep{choi06b,weissman06}.  The apparent motion is
too great for a bound orbit, given the expected mass of Echeclus. Also,
Echeclus's Hill sphere radius at the time was roughly only $5\times10^4$ km,
i.e. about 6 arcsec on the sky.
Further evidence against the satellite hypothesis is that a search
in earlier deep imaging (when no activity was seen) yielded no such
object \citep{rou08}.

The fragment itself could be a few kilometers in radius \citep{rou08}
-- compared to the $\sim$40 km of the primary --  and so a large
impulse would be needed for that much mass to be accelerated up to
the $\sim$15 to 30 m/s required to reach escape velocity. While
typical separation speeds are an order of magnitude smaller
\citep{boehnh04}, we note that a fragment that is flung off
the surface of Echeclus due
to the primary's rotation could be given such a speed,
since in that case the fragment's speed would be proportional to
the primary's radius. 
On the other hand, current evidence
suggests that Echeclus has a relatively long rotation period of 27 hours
\citep{rou05}.

While a fragment remains the best explanation for 174P, further
study of this enigmatic comet is certainly warranted. The main
science goals would be to (a) ascertain the nature of Echeclus's
activity so as to explain how such a large mass could have left the
primary; (b) explain why activity is so tightly localized; (c)
determine what the source of the activity is (CO? CO$_2$?
crystallization of amorphous H$_2$O ice?) and whether this plays a
role in making fragmentation more likely; (d) monitor the rest of
the Centaur population to determine whether any other objects suffer
these events, and what the frequency is; and (e) infer what Centaur
fragmentation implies for the bulk structure and strength of the
JFCs that the Centaurs become.

\subsection{Short-Period SOHO Comets}

The SOHO spacecraft has discovered over 1500 comets, almost all of
which are ``sungrazers" (comets with perihelia less than about 0.06
AU).  A large fraction of the SOHO comets are part of the Kreutz
family, whose comets pass only about 0.005 AU from the center of
the Sun.  Since the Sun's radius is 0.00465 AU, many Kreutz comets
(and almost all the ones discovered by SOHO, which are faint)
disintegrate to dust very near perihelion if not earlier.  The
Kreutz comets are all presumably  fragments of a long-period comet
that broke apart thousands of years ago; see \citet{mars05a} for a
recent review.

Since the Kreutz comets are of long-period, we do not consider them
here. However SOHO has discovered four other families of sungrazer
comets, some of which may be of short-period. While the Meyer group
\citep{meyer02} comets have orbital inclinations of about 72$^\circ$,
well out of the ecliptic, the Marsden group \citep{mars02b}, the
Kracht group \citep{kracht02a}, and the Kracht II group \citep{kracht02b}
have members with inclinations of 27$^\circ$, 13$^\circ$, and
13$^\circ$, respectively, perfectly normal for short-period comets.
As with the Kreutz family, the members of each group have similar
orbits and thus imply a single progenitor in the past. The perihelia
of these comets are at about 0.05 AU, so, while not currently passing
within the Sun's Roche limit, perhaps stress due to tidal forces 
or due to energy transport
facilitated fragmentation at higher heliocentric distances.
Such distant fragmentation
has been suggested by \citet{sek02} for the Kreutz
comets.

Recent work incorporating the 12-year database of cometary astrometry
from SOHO has revealed possible linkages between some of the comets
in these three groups, as well as between some comets
not belonging to any known group. This information is collected in Table 2.
It is worth noting that the linking of apparitions with short-arc
orbits can be difficult, especially when independent comets move
on similar orbits anyway.  There are several other possible links
among comets within the three groups that have been reported
\citep[e.g.][]{mars06,kracht08}. \citet{sc05} present a detailed
analysis of some of the Marsden and Kracht group linkages.

Detailed physical studies of the sungrazing short-period comets are
currently limited to observations by Sun-staring spacecraft. These
often provide magnitudes and thus some secular light curve. In many
cases, the  comets are sufficiently faint that neither a tail nor
an extended coma is visible, limiting the amount of available
information on gas and dust.  The nuclei are probably of order
dekameters in radius \citep{mars05a}; this means that even with an
improved orbit, it would be feasible to observe such a comet well
away from the Sun only when it is near Earth. For example, a 10-meter
radius nucleus at opposition 0.1 AU from Earth still only achieves
an R-band magnitude of 22.5. So even in a very favorable apparition (which
would not happen very often to begin with),
it will not be easy to make detailed studies of such comets. 
The detailed study of a
wide sample of sungrazing short-period comets will probably have
to wait for very deep and wide sky surveys, or for classically-scheduled
time at the largest telescopes.

\subsection{Cometary Debris}

As comet 73P shows, comet fragments have a distribution of sizes.
Indeed, Fig. 1 demonstrates that such a comet gives off fragments
that are decimeter sized and larger -- fragments that are larger
than what can be typically lifted off a comet's surface simply by
gas drag \citep{gj90}.  These fragments remain in a trail in the
orbit plane since radiation pressure acts slowly on them.

The deep infrared observations by IRAS brought to light the existence
of cometary debris trails \citep{sw92}, consisting mainly of
millimeter and  centimeter sized grains that are the largest solid
bodies that come off the comet during ``normal" activity.  Such
grains have often been found to contain most of the mass that is
contained in dust.  The grains in the trails represent an intermediate
size scale between the visible-wavelength dust that is typically
micron sized and smaller, and observed fragments that are dekameter
sized and larger.

An infrared survey of debris trails has been performed by Spitzer.
\citet{reach07a} present 24 $\mu$m imaging of 34 JFCs in which a
trail is unambiguously seen in 27 of them -- about 80\%.  Most of
these comets were observed while 2 to 4 AU from the Sun, and months
away from perihelion. Examples of the
observations presented by \citet{reach07a} are shown in Fig. 6.

Most of the comets in their survey had not suffered splitting
events in the traditional sense; the particles were
liberated by regular cometary activity. One could call
these millimeter and centimeter sized particles ``fragments"
only in a liberal definition.  Nonetheless, the
high frequency of trails, coupled with the fact that often the trail
was seen all the way out to the edge of the field of view, indicates
that JFCs are prodigious producers of millimeter and
centimeter sized fragments into 
the interplanetary dust environment. As \citet{reach07a}
point out, the total mass of these large grains is larger
than the total mass of grains seen at visible wavelengths.

\subsection{Searches for Fragments}

Usually fragments are found serendipitously, but a few pointed
searches for fragments have been performed.  The survey of \citet{cj94},
mentioned earlier, focussed on finding fragments near (i.e. a few
arcminutes away from) the primary comet. The faint meter and dekameter
scale objects could have been missed, however.  \citet{beech04} did
a telescopic search for meter-sized (and larger) Perseid meteoroids
while Earth was passing through the meteor stream; this would have
detected fragments of the Halley-family comet 109P/Swift-Tuttle.
They sought to find objects a few arcminutes away from the radiant
before they collided with Earth's atmosphere. The search found no
objects, thus constraining the space density of meter-scale objects
within the stream.

Another survey has made use of the proliferation of CCD cameras
with very wide fields-of-view.  \citet{sj07} have used the Megacam
instrument on the CFH Telescope atop Mauna Kea to search for fragments
up to half a degree away from a comet's head. They have observed
about 12 JFCs so far with no fragments seen to a limiting magnitude
of about 24. For these comets this limit corresponds to fragments
of size about 100 m in radius.  With such a large field-of-view,
the search is not just a snapshot in time, but actually tests whether
these comets have had any splitting event in the previous several
months. This kind of survey gives a more complete view of the
fragmentation history and so should provide a good new measurement
of the frequency of JFC splittings.

\section{Concluding Remarks}

The interested reader is encouraged to study the recent review of
split comets given by \citet{boehnh04}. We have tried here to provide
a summary of recent developments on the topic of split comets. But
what does the future hold for our understanding of this phenomenon?
While splittings caused by tides can be predicted if the comet's orbit is
known, the large majority of JFC splittings instead happen
stochastically. Classically-scheduled observing runs will often
fail to catch these events. Synoptic observations of comets however
will in general be more fruitful for understanding the full evolution
of a comet before, during, and after a fragmentation event; unbiassed
and well-sampled datasets will be important.

It is possible that all-sky survey projects  in the coming decade
will indeed provide us with a great deal of such data.  For example,
Pan-STARRS and LSST will likely discover several hundred comets
each in their first year of operation in addition to the thousands
of new Trojans, Centaurs, and trans-Neptunian objects they will
find \citep{jew03, ive08}.  In addition, the temporal coverage
(scanning the sky on week-long timescales), depth (reaching
approximately 23rd magnitude), and angular resolution will ensure
that moderate-sized fragments of JFCs will be found more reliably.
Furthermore, we will be able to follow the photometric and dynamic
evolution of the fragments for months and possibly years at a time.
The datasets will revolutionize our picture of fragment behavior
and endurance  (in addition to revolutionizing many questions about
comets).

Since the review by \citet{boehnh04} was published, the most important
event with regard to split JFCs was the apparition of comet
73P/Schwass\-mann-Wachmann 3 in 2006.  Following up other known
split comets in the future would be worthwhile as well but 73P
represents a unique case that is crucial to study.  Observations
of 73P in  forthcoming years can give us new details about its
fragments and its continuing evolution. If fragments B and C survive,
we will be able to watch their new surfaces age as they continue
to be exposed to the space environment. Will the resulting chemistry
change the apparent production rates we measure in their comae? If
so, how? Observations of other JFCs at multiple apparitions have
shown a constancy in relative coma abundances from orbit to orbit
\citep{ahe95}, but likely none of those comets have had such a
recent fragmentation event that produced a secondary of comparable
size to the primary.  Physical observations of the continued
unraveling of fragment B and perhaps the eventual unraveling of
fragment C would also give us more insight into cometary structure.
Unfortunately, three of the next five apparitions of 73P are
unfavorable -- only the visits in 2022 and 2033 are good -- but
even limited data on each fragment at each orbit would be useful.

More generally, wide-field observations of all JFCs
that will make close approaches to Earth would be useful for
understanding the frequency of fragmentation and
the endurance of the fragments, since the
proximity will make searches for meter-scale fragments
feasible.  For example, 
comet 103P/Hartley 2 approaches to 0.12 AU in 2010, and comet
45P/Honda-Mrkos-Pajdu\v sakov\'a comes to 0.06 AU in
2011.

\ack{We thank W. T. Reach, J. M. Bauer, H. A. Weaver, and
T. Fuse for the use of their figures, and M. S. Kelley
for helpful suggestions. We appreciate the
thorough reviews of this manuscript given by N. Samarasinha
and an anonymous referee.}


\vfill
\eject

\begin{table}
\begin{center}
\caption{Known Split JFCs \vskip 5 pt}
\begin{tabular}{ll}
\hline
Comet & When \\
\hline
3D/Biela & 1840 \\
16P/Brooks 2  & 1889, 1995 \\
51P/Harrington & 1994, 2001 \\
57P/du Toit-Neujmin-Delporte & 2002 \\
69P/Taylor & 1915 \\
73P/Schwassmann-Wachmann 3 & 1995, 2001, 2006 \\
79P/du Toit-Hartley & 1982 \\
101P/Chernykh & 1991, 2005 \\
108P/Ciffreo & 1985 \\
120P/Shoemaker-Holt 1 & 1996 \\
141P/Machholz 2 & 1987, 1989 \\
174P/Echeclus  & 2006 \\
205P/Giacobini & 1896, 2008 \\
D/1993 F2 Shoemaker-Levy 9 & 1992 \\
P/2004 V5 (LINEAR-Hill) & 2004 \\ 
\hline
\noalign{\vskip 5pt}
\multispan2{List adapted from work by \citet{boehnh04}. \hfil} \\
\multispan2{Note that 174P is also a Centaur. \hfil }\\ 
\noalign{\vskip 15pt}
\end{tabular}
\end{center}
\end{table}

\begin{table}
\begin{center}
\caption{Potentially Periodic Sungrazers \vskip 5 pt}
\begin{tabular}{lcrc}
\hline
Comets & Group & $P$ &  Refs. \\
\hline
C/1999 J6 = C/2004 V9  & Marsden & 5.49 & 1 \\
C/1999 M3 = C/2004 L10 & Kracht & 4.95 & 1,2 \\
C/1999 N5 = C/2005 E4 & Marsden &  5.66 & 3 \\
C/1999 N6 = C/2004 J4 or C/2004 J18 & Kracht & 4.81 & 1,2 \\
C/1999 R1 = C/2003 R5 = C/2007 R5 & Kracht II & 3.99 & 4 \\
C/1999 X3 = C/2004 E2 = C/2008 K10 & (none) & 4.22 & 5 \\
C/2000 O3 = C/2005 W4 & Kracht & 5.32 & 6  \\
C/2001 D1 = C/2004 X7 = C/2008 S2 &  (none) & 3.78 & 7 \\
C/2002 Q8 = C/2008 E4 & Kracht & 5.52 & 8 \\
C/2002 R1 = C/2008 A3   & Marsden & 5.37  & 9 \\
C/2002 S11 = C/2008 G6 & Kracht &  5.54 & 10 \\
\hline
\noalign{\vskip 5pt}
\multispan4{All comets are named SOHO. $P = $ 
orbital period in years. } \\
\multispan4{References: 1 = \citet{mars04}, 2 = \citet{sc05}, }\\
\multispan4{3 = \citet{mars05b}, 4 = \citet{mars07}, 5 = \citet{mars08d}, } \\
\multispan4{6 = \citet{mars05c}, 7 = \citet{mars08e}, 8 = \citet{mars08c},  }\\
\multispan4{9 = \citet{mars08a}, 10 = \citet{mars08b}.  \hfil }\\
\end{tabular}
\end{center}
\end{table}

\newpage

\newpage

\newpage

\newpage

\newpage

\begin{figure}
\includegraphics[width=5.5in]{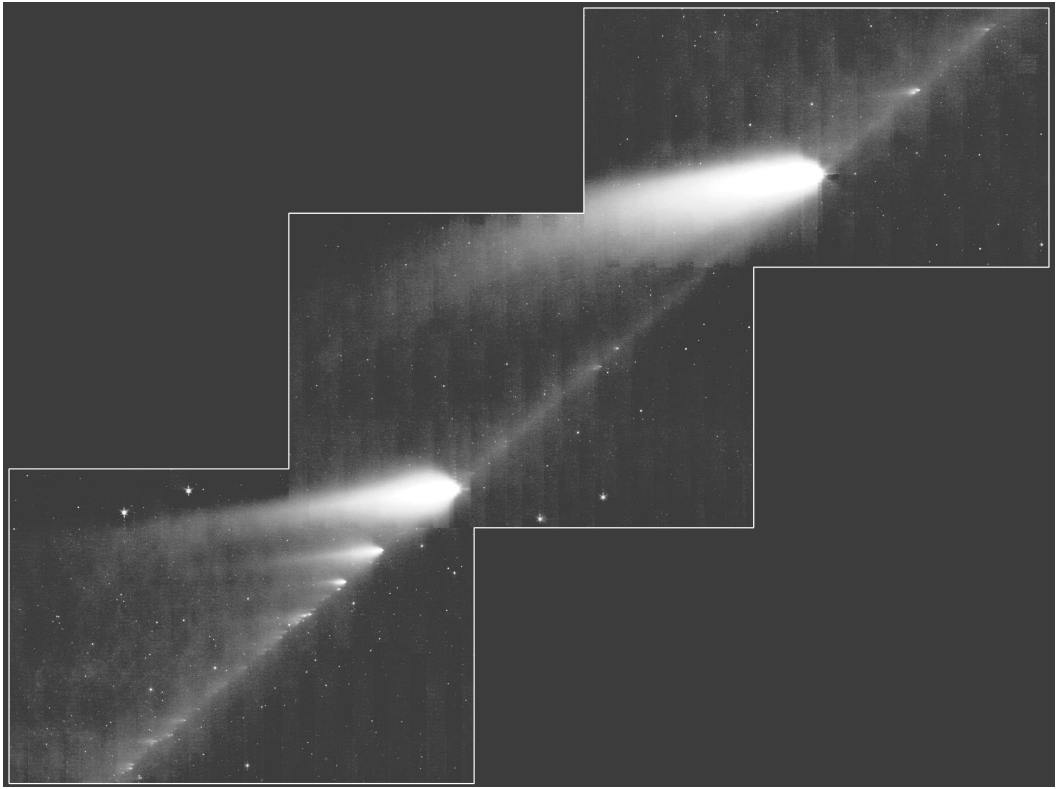}
\caption{
Mosaic of comet 73P/Schwassmann-Wachmann 3
as observed by the Spitzer Space Telescope over May 4 to 6, 
2006, at a wavelength of 24 $\mu$m. About three dozen fragments 
of this split comet are
visible here, and almost every one has its own cometary tail. The fragments
themselves all lie on the comet's projected orbit. Courtesy W. T. Reach
of Caltech.
\hfil \break \vbox to 1cm{\ }}
\end{figure}

\medskip

\begin{figure}
\includegraphics[width=5.5in]{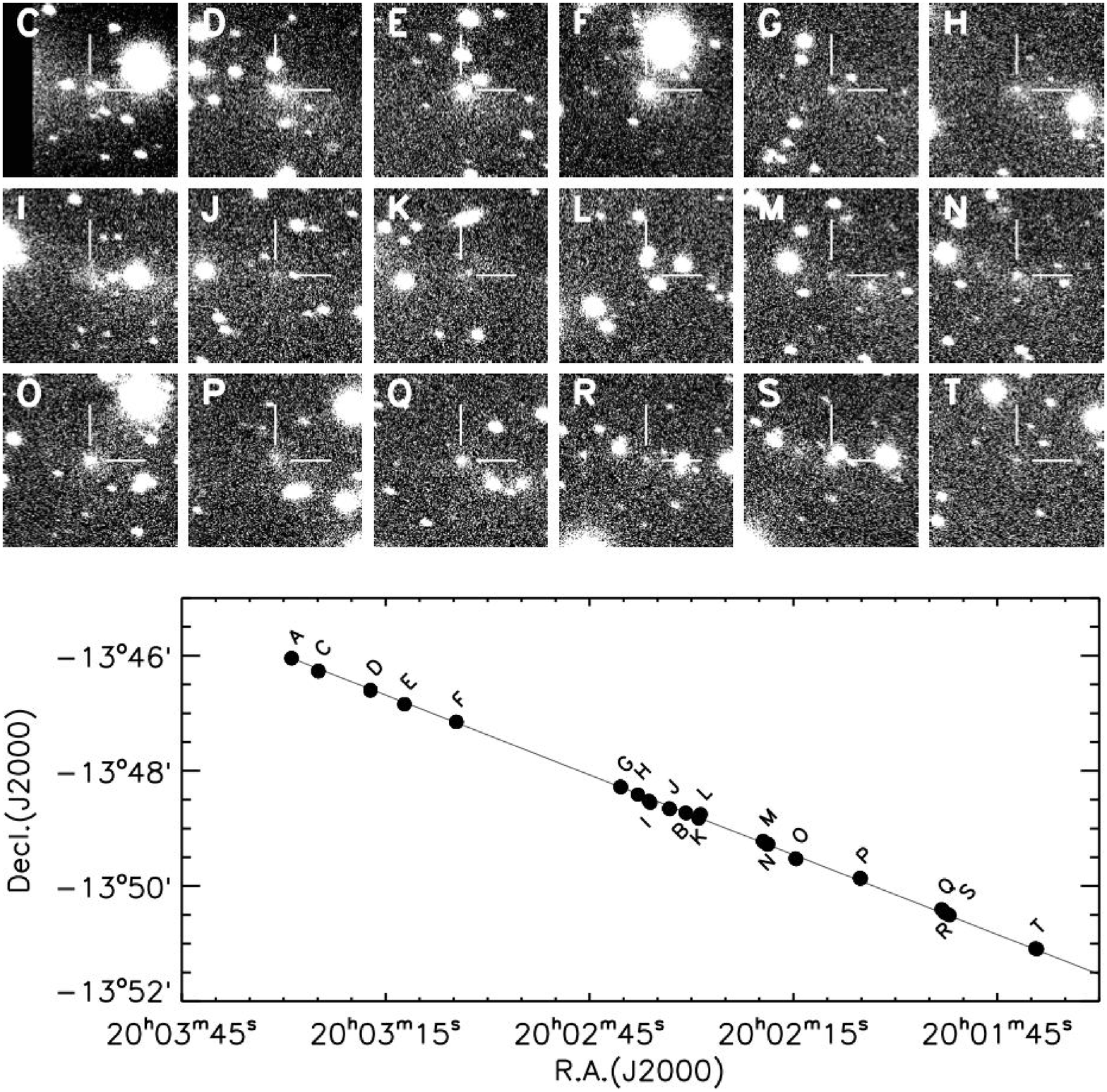}
\caption{
Information about the fragments of 57P. At top are
the 18 fragments 
discovered by \citet{fern02} on July 17, 2002. Each panel is 44 arcsec
across. Short white segments indicate the location of
each fragment. Note the wide variety of morphologies
and condensations. At bottom is a schematic showing the
location of each fragment with respect to the head - fragment A.
The location of fragment B \citep{mars02a} is also shown.
\hfil \break \vbox to 1cm{\ }}
\end{figure}

\medskip

\begin{figure}
\includegraphics[width=5.5in]{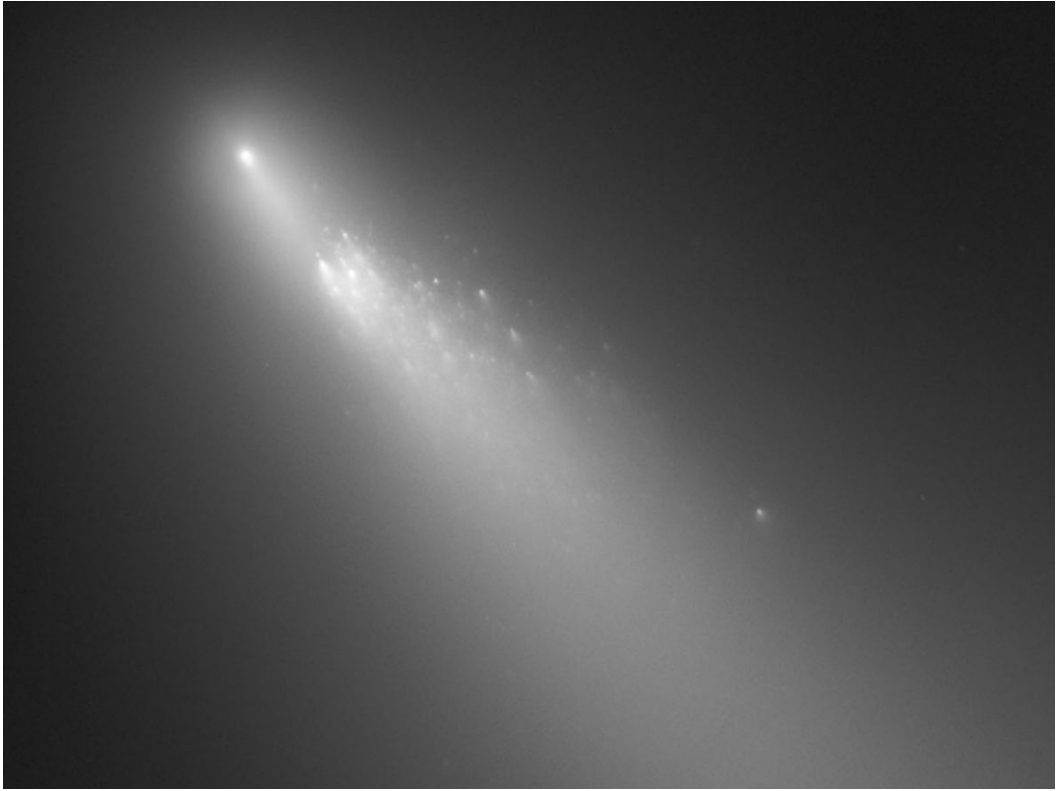}
\caption{
Hubble Space Telescope image of
fragment B of comet 73P, taken on April 18, 2006
by \citet{weaver06}. The
fragment itself is the condensation at upper left;
the image was taken just after the fragment had
shed several fragments of its own. These are
the dozens of condensations farther down the tail.
The image is about 25 arcsec across, and
HST's
spatial resolution is just 8 km/pixel (i.e. 0.05 arcsec/pixel).
Courtesy H. A. Weaver of JHU APL.  
\hfil \break \vbox to 1cm{\ }}
\end{figure}

\medskip

\begin{figure}
\includegraphics[width=5.5in]{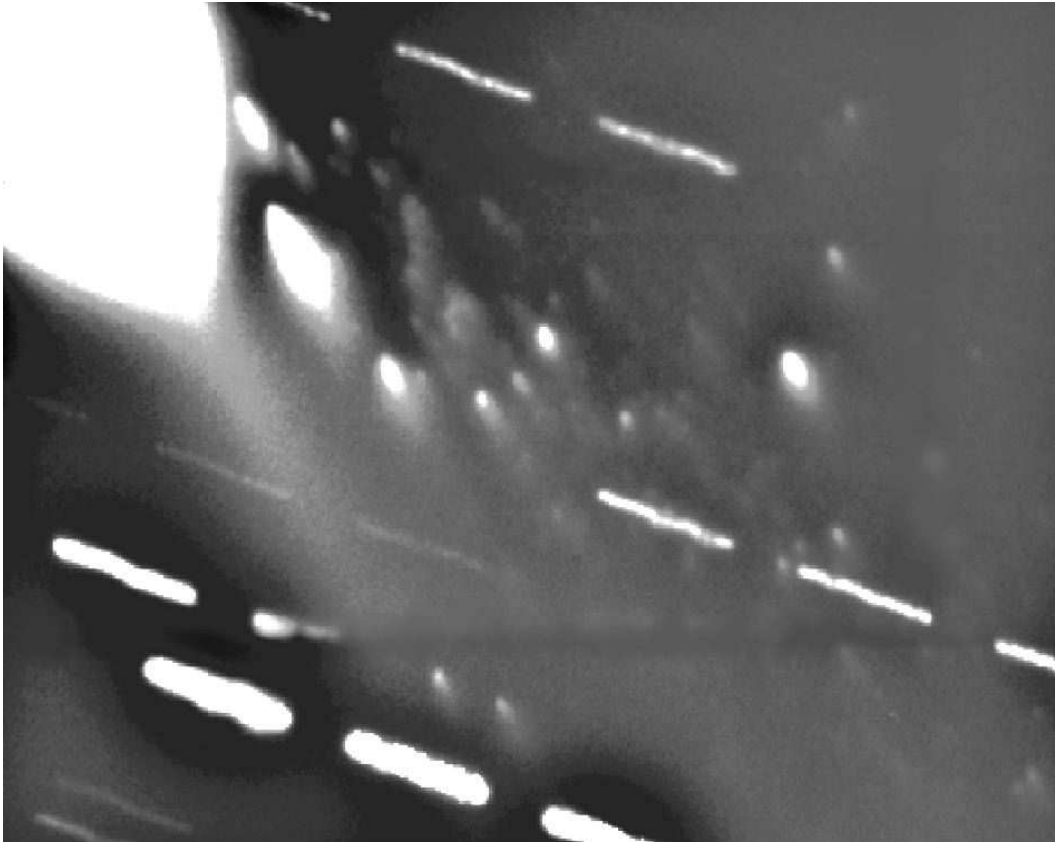}
\caption{
Detail of an image (after unsharp masking)
of fragment B taken by \citet{fuse07}
on May 3, 2006, with the Subaru telescope. The image is their
Figure 1. The field 
size is 96 by 68 arcsec. 
Fragment B itself is in the upper left, and 
\citet{fuse07} report finding 54 fragments in the image. 
Compare the time and scale to Fig. 3. White streaks are trailed stars. 
\hfil \break \vbox to 1cm{\ }}
\end{figure}

\medskip

\begin{figure}
\includegraphics[width=5.5in]{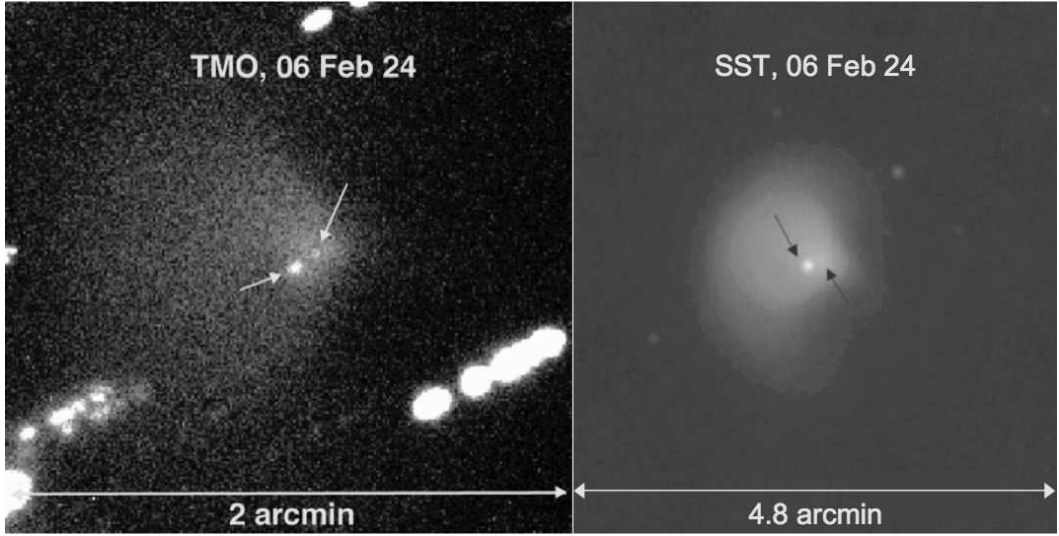}
\caption{
Images of comet 174P from February 24, 2006,
adapted from Figs. 1 and 3 in the work by \citet{bauer08}. The left panel
shows the R-band image from the Table Mountain Observatory
0.6-m telescope; the right shows the 24-$\mu$m image from
the Spitzer Space Telescope. In each panel, the two arrows
indicate the main body of Echeclus itself and the condensation
of the coma. Note that the orientations are slightly different; in the
TMO image, equatorial north is up, while in
the SST image, north is 21$^\circ$ to the right of up.
\hfil \break \vbox to 1cm{\ }}
\end{figure}

\medskip

\begin{figure}
\includegraphics[width=5.5in]{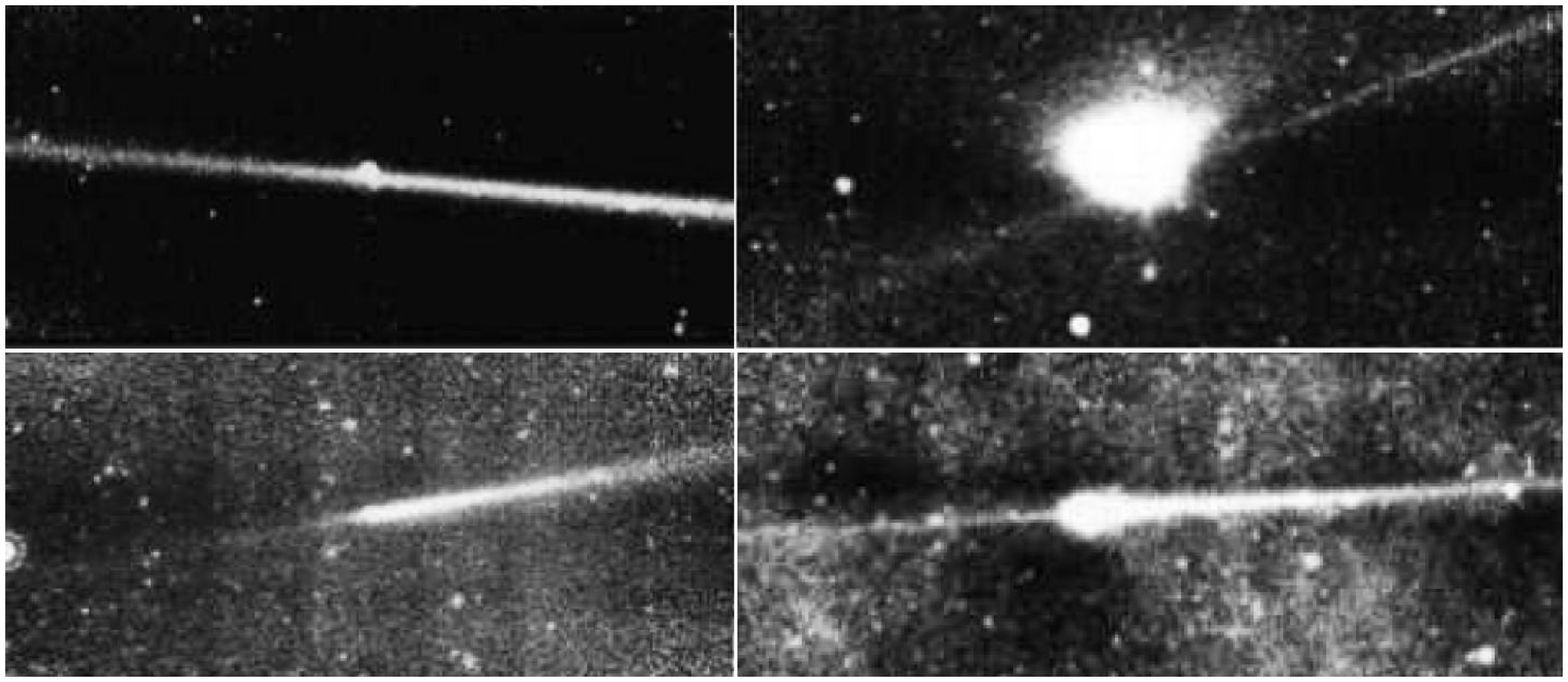}
\caption{
Spitzer images of four representative JFCs taken from
the work by \citet{reach07a}. Clockwise from top-left: 10P/Tempel 2,
48P/Johnson, 129P/Shoemaker-Levy 3, and 
67P/Churyumov-Gerasimenko. All show
a linear trail populated by millimeter (and larger) scale grains, 
though with different
meridional and longitudinal variations. While some of the comet-to-comet
variation is due to observing geometry, there are some intrinsic differences
in the trails. 
\hfil \break \vbox to 1cm{\ }}
\end{figure}


\begin{thebibliography}{}


%

\bibitem[A'Hearn et al.(1995)]{ahe95} A'Hearn, M.F., Millis, R.L.,
Schleicher, D.G., Osip, D.J., Birch, P.V., 1995.  The ensemble
properties of comets: Results from narrowband photometry of 85
comets, 1976-1992.  Icarus 118, 223--270.

\bibitem[A'Hearn(2008)]{ah08} A'Hearn, M.F., 2008. Deep Impact and
the origin and evolution of cometary nuclei. Space Sci. Rev., 138, 237--246.

\bibitem[Bauer et al.(2008)]{bauer08} Bauer, J.M., Choi, Y.-J.,
Weissman, P.R., Stansberry, J.A., Fernandez, Y.R., Roe, H.G.,
Buratti, B.J., Sung, H.-I., 2008. The large-grained dust coma of
174P/Echeclus.  Publ. Astron. Soc. Pac. 120, 393--404.

\bibitem[Beech et al.(2004)]{beech04} Beech, M., Illingworth, A.,
Brown, P., 2004. A telescopic search for large Perseid meteoroids.
Mon. Not.  Roy. Astron. Soc. 348, 1395--1400.

\bibitem[Beech and Nikolova(2001)]{bn01} Beech, M., Nikolova, S.,
2001. The endurance lifetime of ice fragments in cometary streams.
Plan. \& Space Sci. 49, 23--29.

\bibitem[Birtwhistle(2008)]{birt08} Birtwhistle, P., 2008.
73P/Schwassmann-Wachmann (fragment summary 2006). URL {\tt
http://www.birtwhistle.org/Gallery73P-2006Summary.htm}.  Accessed
June 25, 2008.

\bibitem[Biver et al.(2006)]{biv06} Biver, N., Bockel\'ee-Morvan,
D., Boissier, J., Colom, P., Crovisier, J., Lecacheux, A., Lis, D.
C., Parise, B., Menten, K., and the Odin team, 2006.  Comparison
of the chemical composition of fragments B and C of comet
73P/Schwassmann-Wachmann 3 from radio observations. Bull. Amer.
Astron. Soc. 38, 484. 

\bibitem[Boehnhardt(2002)]{boehnh02} Boehnhardt, H., 2002. Comet
splitting observations and model scenarios. Earth Moon \& Plan. 89,
91--115.

\bibitem[Boehnhardt(2004)]{boehnh04} Boehnhardt, H., 2004. Split
comets.  In: Festou, M. C., Keller, H. U., Weaver, H. A. (Eds.),
Comets II, pp. 301--316.  University of Arizona Press.

\bibitem[Boehnhardt et al.(2002)]{betal02} Boehnhardt, H., Holdstock,
S., Hainaut, O., Tozzi, G.P., Benetti, S., Licandro, J., 2002.
73P/Schwassmann-Wachmann 3 - One orbit after breakup: Search for
fragments.  Earth Moon \& Plan. 90, 131--139.

\bibitem[Boehnhardt et al.(1999)]{boehnh99} Boehnhardt, H., Rainer,
N., Birkle, K., Schwehm, G., 1999. The nuclei of comets
26P/Grigg-Skjellerup and 73P/Schwassmann-Wachmann 3. Astron. \&
Astrophys. 341, 912--917.

\bibitem[Bonev et al.(2008)]{bonev08} Bonev, T., Boehnhardt, H.,
and Borisov, G., 2008. Broadband imaging and narrowband polarimetry
of comet 73P/Schwassmann-Wachmann 3, components B and C, on 3, 4,
8, and 9 May 2006. Astron. \& Astrophys. 480, 277--287.

\bibitem[Carusi et al.(1985)]{car85} Carusi, A., Perozzi, E.,
Valsecchi, G.B., Kresak, L., 1985. First results of the integration
of motion of short-period comets over 800 years. In: Carusi, A.,
Valsecchi, G.B. (Eds.), Dynamics of Comets: Their Origin and
Evolution, pp. 319--340.  D. Reidel Publishing Co.

\bibitem[Chen and Jewitt(1994)]{cj94} Chen, J., Jewitt, D., 1994.
On the rate at which comets split.  Icarus 108, 265--271.

\bibitem[Choi et al.(2006a)]{choi06a} Choi, Y.-J., Weissman, P. R.,
Polishook, D., 2006a. (60558) 2000 EC$_{98}$. IAU Circ. 8656.

\bibitem[Choi et al.(2006b)]{choi06b} Choi, Y.-J., Weissman, P.R.,
Chesley, S., Bauer, J., Stansberry, J., Tegler, S., Romanishin, W.,
Consolmagno, G., 2006b. Comet 174P/Echeclus.  Central Bureau Electr.
Teleg. 563.

\bibitem[Dello Russo et al.(2007)]{ndr07} Dello Russo, N., Vervack,
R.J., Weaver, H.A., Biver, N., Bockel\'ee-Morvan, D., Croivisier,
J., Lisse, C.M., 2007. Compositional homogeneity in the fragmented
comet 73P/Schwassmann-Wachmann 3.  Nature 448, 172--175.

\bibitem[DiSanti et al.(2007)]{dis07} DiSanti, M. A., Anderson,
W.M., Villanueva, G.L., Bonev, B.P., Magee-Sauer, K., Gibb, E.L.,
Mumma, M.J., 2007. Depleted carbon monoxide in fragment C of the
Jupiter-Family comet 73P/Schwassmann-Wachmann 3.  Astrophys. J.
661, L101-L104.

\bibitem[DiSanti and Mumma(2008)]{dm08} DiSanti, M. A., Mumma, M.
J., 2008. Reservoirs for comets: Compositional differences based
on infrared observations. Space Sci. Rev., 138, 127--145.

\bibitem[Ernst and Schultz(2007)]{es07} Ernst, C. M., Schultz, P.
H., 2007. Evolution of the Deep Impact flash: Implications for the
nucleus surface based on laboratory experiments. Icarus 190, 334--344.

\bibitem[Feaga et al.(2007)]{fea07} Feaga, L. M., A'Hearn, M.F.,
Sunshine, J.M., Groussin, O., Farnham, T.L., 2007.  Asymmetries in
the distribution of H$_2$O and CO$_2$ in the inner coma of comet
9P/Tempel 1 as observed by Deep Impact.  Icarus 190, 345--356.

\bibitem[Fernandez et al.(2002)]{fern02} Fernandez, Y. R., Jewitt,
D. C., Sheppard, S. S., 2002. Comet 57P/du Toit-Neujmin-Delporte.
IAU Circ. 7935.

\bibitem[Fuse et al.(2007)]{fuse07} Fuse, T., Yamamoto, N., Kinoshita,
D., Furusawa, H., Watanabe, J., 2007. Observations of fragments
split from nucleus B of comet 73P/Schwassmann-Wachamnn 3 with Subaru
Telescope. Publ. Astron. Soc. Japan 59, 381--386.

\bibitem[Groussin et al.(2007)]{grou07} Groussin, O., A'Hearn, M.F.,
Li, J.-Y., Thomas, P.C., Sunshine, J.M., Lisse, C.M., Meech, K.J.,
Farnham, T.L., Feaga, L.M., Delamere, W.A., 2007.  Surface temperature
of the nucleus of comet 9P/Tempel 1.  Icarus 187, 16--25.

\bibitem[Gr\"un and Jessberger(1990)]{gj90} Gr\"un, E., Jessberger,
E. K., 1990. Dust. In: Huebner, W. F. (Ed.),  Physics and Chemistry
of Comets, pp. 113--175. Springer-Verlag.

\bibitem[Hanner et al.(1981)]{hann81} Hanner, M. S., Giese, R. H.,
Wiess, K., Zerull, R., 1981. On the definition of albedo and
application to irregular partcles. Astron. \& Astrophys. 104, 42--46.

\bibitem[Holsapple and Housen(2007)]{hh07} Holsapple, K. A., Hausen,
K. R., 2007. A crater and its ejecta: An interpretation of Deep
Impact. Icarus 187, 345--356.

\bibitem[Howell et al.(2007)]{how07} Howell, E. S., Nolan, M.C.,
Harmon, J.K., Lovell, A.J., Benner, L.A., Ostro, S.J., Campbell,
D.B., Margot, J., 2007. Radar and radio observations of the fragmented
comet 73P/Schwassmann-Wachmann 3.  Bull. Amer. Astron. Soc. 39,
486. 

\bibitem[Hughes and McBride(1992)]{hm92} Hughes, D. W., McBride,
N., 1992. Short-period comet splitting. J. Brit. Astron. Assoc.
102, 265--268.

\bibitem[Ivezi\'c et al.(2008)]{ive08} Ivezi\'c, \v Z., and 93
colleagues, 2008. LSST: From science drivers to reference design
and anticipated data products. arXiv:0805.2366v1 [astro-ph] 15 May
2008.

\bibitem[Jewitt(2003)]{jew03} Jewitt, D., 2003. Project Pan-STARRS
and the outer Solar System. Earth Moon \& Plan. 92, 465--476.

\bibitem[Jewitt(1990)]{jewitt90} Jewitt, D., 1990. 
The persistent coma of comet P/Schwassmann-Wachmann 1.
Astrophys. J. 351, 277--286.

\bibitem[Kobayashi et al.(2007)]{kob07} Kobayashi, H., Kawakita,
H., Mumma, M.J., Bonev, B.P., Watanabe, J., Fuse, T., 2007.  Organic
volatiles in comet 73P-B/Schwassmann-Wachmann 3 observed during its
outburst: A clue to the formation region of the Jupiter-Family
comets. Astrophys. J. 668, L75--L78.

\bibitem[Kracht(2002a)]{kracht02a} Kracht, R., 2002a. Comets C/1999
M3, 2002 E1 (SOHO). Minor Plan. Electr. Circ. 2002-E18.

\bibitem[Kracht(2002b)]{kracht02b} Kracht, R., 2002b.  Comets C/2002
R4, 2002 R5, 2002 R6, 2002 R7, 2002 R8 (SOHO).  Minor Plan. Electr.
Circ. 2002-S35.

\bibitem[Kracht(2008)]{kracht08} Kracht, R., 2008. Orbital links
between the Marsden and between the Kracht group comets. URL {\tt
http://www.rkracht.de/soho/links.htm}. Accessed July 3, 2008.

\bibitem[Lien(1990)]{lien90} Lien, D. J., 1991. Optical properties
of cometary dust.  In: Newburn, R.L., Neugebauer, M., Rahe, J.
(Eds.),  Comets in the post-Halley era, pp. 1005--1041.  Kluwer.

\bibitem[Levison and Duncan(1997)]{ld97} Levison, H. F., Duncan,
M. J., 1997. From the Kuiper Belt to Jupiter-Family comets: The
spatial distribution of ecliptic comets. Icarus 127, 13--32.

\bibitem[Lorin and Rousselot(2007)]{lr07} Lorin, O., Rousselot, P.,
2007.  Search for cometary activity in three Centaurs [(60558)
Echeclus, 2000 FZ$_{53}$ and 2000 GM137] and two trans-Neptunian
objects [(29981) 1999 TD$_{10}$ and (28978) Ixion]. Mon. Not. Roy.
Astron. Soc. 376, 881--889.

\bibitem[Lowry and Fitzsimmons(2001)]{lf01} Lowry, S. C., Fitzsimmons,
A., 2001. CCD photometry of distant comets II. Astron. \& Astrophys.
365, 204--213.

\bibitem[Marsden(2002a)]{mars02a} Marsden, B. G., 2002. Comet 57P/du
Toit-Neujmin-Delporte. IAU Circ.  7934.

\bibitem[Marsden(2002b)]{mars02b} Marsden, B. G., 2002. Comet C/2002
C3 (SOHO).  Minor Plan. Electr. Circ.  2002-C28.

\bibitem[Marsden(2004)]{mars04} Marsden, B. G., 2004. Comets C/2004
V9, 2004 V10 (SOHO).  Minor Plan. Electr. Circ.  2004-X73.

\bibitem[Marsden(2005a)]{mars05a} Marsden, B. G., 2005a.  Sungrazing
comets. Ann. Rev. Astron. \& Astrophys. 43, 75--102.

\bibitem[Marsden(2005b)]{mars05b} Marsden, B. G., 2005b. Comets
C/2005 E3, 2005 E4 (SOHO).  Minor Plan. Electr. Circ. 2005-E87.

\bibitem[Marsden(2005c)]{mars05c} Marsden, B. G., 2005c. Comets
C/2005 W4, 2005 W5 (SOHO).  Minor Plan. Electr. Circ. 2005-X14.

\bibitem[Marsden(2006)]{mars06} Marsden, B. G., 2006.  Comets C/1996
X3, 1996 X4, 1996 X5, 1997 B5, 1997 B6, 1997 B7 (SOHO).  Minor Plan.
Electr. Circ. 2006-C49.

\bibitem[Marsden(2007)]{mars07} Marsden, B. G., 2007.  Comet P/1999
R1 = 2003 R5 = 2007 R5 (SOHO).  Minor Plan. Electr. Circ.  2007-S16.

\bibitem[Marsden(2008a)]{mars08a} Marsden, B. G., 2008a.  Comets
C/2007 Y8, 2007 Y9, 2007 Y10, 2008 A3 (SOHO).  Minor Plan. Electr.
Circ.  2008-B61.

\bibitem[Marsden(2008b)]{mars08b} Marsden, B. G., 2008b.  Comets
C/2008 G5, 2008 G6, 2008 H2, 2008 H3 (SOHO).  Minor Plan. Electr.
Circ. 2008-L29.

\bibitem[Marsden(2008c)]{mars08c} Marsden, B. G., 2008c.  Comets
C/2002 Q8, 2008 E4, 2008 F1 (SOHO).  Minor Plan. Electr. Circ.
2008-F32.

\bibitem[Marsden(2008d)]{mars08d} Marsden, B. G., 2008d.  Comet 
C/1999 X3 = 2004 E2 = 2008 K10 (SOHO).  Minor Plan. Electr. Circ.
2008-S49.

\bibitem[Marsden(2008e)]{mars08e} Marsden, B. G., 2008e.  Comets
C/2001 D1 = 2004 X7 = 2008 S2 (SOHO).  Minor Plan. Electr. Circ.
2008-S82.

\bibitem[Meech et al.(2004)]{mhm04} Meech, K. J., Hainaut, O. R.,
Marsden, B. G., 2004. Comets nucleus size distributions from HST
and Keck telescopes. Icarus 170, 463--491.

\bibitem[Meech and Svore\v n(2004)]{ms04} Meech, K. J., Svore\v n,
J., 2004. Using cometary activity to trace the physical and chemical
evolution of cometary nuclei. In: Festou, M. C., Keller, H. U.,
Weaver, H. A. (Eds.), Comets II, pp. 317--335.  University of Arizona
Press.

\bibitem[Meyer(2002)]{meyer02} Meyer, M., 2002.  Comets C/2002 C3
(SOHO). Minor Plan. Electr. Circ.  2002-C28.

\bibitem[Mumma(2008)]{mumma08} Mumma, M. J., 2008.
Chemical diversity of organic volatiles among comets: An emerging 
taxonomy and implications for processes in the proto-planetary disk.
In: Kwok, S., Sandford, S. A. (Eds.), 
Proceedings IAU Symposium No. 251: Organic Matter in Space,
pp. 309-310. Cambridge University Press.

\bibitem[Reach et al.(2007a)]{reach07a} Reach, W. T., Kelley, M.
S., Sykes, M. V., 2007. A survey of debris trails from short-period
comets. Icarus 191, 298--322.

\bibitem[Reach et al.(2007b)]{reach07b} Reach, W. T., Lisse, C. M.,
Kelley, M. S., Vaubaillon J., 2007. Rocket effect for fragments and
meteoroids of the split comet 73P/Schwassmann-Wachmann 3.  Bull.
Amer. Astron. Soc. 39, 524. 

\bibitem[Rousselot(2008)]{rou08} Rousselot, P., 2008. 174P/Echeclus:
A strange case of outburst. Astron. \& Astrophys. 480, 543--550.

\bibitem[Rousselot et al.(2005)]{rou05} Rousselot, P., Petit, J.-M.,
Poulet, F., Sergeev, A., 2005. Photometric study of Centaur (60558)
2000 EC$_{98}$ and trans-neptunian object (55637) 2002 UX$_{25}$
at different phase angles. Icarus 176, 478--491.

\bibitem[Samarasinha(2007)]{sama07} Samarasinha, N. H., 2007.
Rotation and activity of comets. Adv. Space. Res. 39, 421--427.

\bibitem[Schleicher et al.(2006)]{schl06} Schleicher, D. G., Birch,
P. V., Bair, A. N., 2006. The composition of the interior of comet
73P/Schwassmann-Wachmann 3: Results from narrowband photometry of
multiple components.  Bull. Amer. Astron. Soc. 38, 485. 

\bibitem[Schleicher and Bair(2008)]{sb08} Schleicher, D. G.,
Bair, A. N., 2008. Compositional taxonomy of comets and 
the unique cases of 96P/Machholz 1 and 
73P/Schwassmann-Wachmann 3.  In: LPI Editorial Board (Eds.),
LPI Contribution No. 1405: Asteroids, Comets, Meteors 2008
(CD-ROM), paper ID 8174.

\bibitem[Scotti et al.(2000)]{sco00} Scotti, J. V., Gleason, A. E.,
Montani, J. L., Read, M. T., 2000. Eight TNOs and Centaurs.  Minor
Plan. Electr. Circ.  2000-E64.

\bibitem[Sekanina(1982)]{seka82} Sekanina, Z., 1982. The problem
of split comets in review. In: Wilkening, L. L. (Ed.), Comets, pp.
251--287.  University of Arizona Press.

\bibitem[Sekanina(1997)]{sek97} Sekanina, Z., 1997.  The problem
of split comets revisited. Astron. \& Astrophys. 318, L5--L8.

\bibitem[Sekanina(2002)]{sek02} Sekanina, Z., 2002.  Runaway
fragmentation of sungrazing comets observed with the Solar and
Heliospheric Observatory. Astrophys. J. 576, 1085--1089.

\bibitem[Sekanina(2007)]{sek07} Sekanina, Z., 2007.  Earth's 2006
encounter with comet 73P/Schwassmann-Wachmann: Products of nucleus
fragmentation seen in closeup.  In: Valsecchi, G.B., Vokrouhlicky,
D. (Eds.), Near Earth Objects, our celestial neighbors: Opportunity
and risk, pp. 211--220.  Cambridge University Press.

\bibitem[Sekanina and Chodas(2002)]{sc02} Sekanina, Z., Chodas, P.
W., 2002. Comet 57P/du Toit-Neujmin-Delporte. IAU Circ. 7957.

\bibitem[Sekanina and Chodas(2005)]{sc05} Sekanina, Z., Chodas, P.
W., 2005.  Origin of the Marsden and Kracht groups of sunskirting
comets. I.  Association with comet 96P/Machholz and its interplanetary
complex. Astrophys. J. Supp. Ser. 161, 551--586.

\bibitem[Stansberry et al.(2008)]{stansb08} Stansberry, J., Grundy,
W., Brown, M., Cruikshank, D., Spencer, J., Trilling, D., Margot,
J.-L., 2008. Physical properties of Kuiper Belt and Centaur objects:
Constraints from the Spitzer Space Telescope.  In: Barucci, M. A.,
Boehnhardt, H., Cruikshank, D.P., Morbidelli, A. (Eds.), The Solar
System beyond Neptune, pp. 161--179.  University of Arizona Press.

\bibitem[Stevenson and Jedicke(2007)]{sj07} Stevenson, R., Jedicke,
R. 2007. An observational limit of decameter scale fragment mass
loss from comets.  Bull. Amer. Astron. Soc. 39, 525. 
54.09

\bibitem[Storm et al.(2006)]{storm06} Storm, S., Samarasinha, N.,
Mueller, B., Farnham, T., Fernandez, Y., Kidder, A., Snowden, D.,
A'Hearn, M., Harris, W., Knight, M., Morgenthaler, J., Lisse, C.,
Roesler, F., 2006.  Time variability of component C of the fragmented
comet 73P/Schwassmann-Wachmann 3.  Bull. Amer. Astron. Soc.  38,
504. 

\bibitem[Sykes and Walker(1992)]{sw92} Sykes, M. V., Walker, R. G.,
1992. Cometary dust trails. I - Survey. Icarus 95, 180--210.

\bibitem[Sunshine et al.(2007)]{sun07} Sunshine, J. M., Groussin,
O., Schultz, P.H., A'Hearn, M.F., Feaga, L.M., Farnham, T.L.,
Klaasen, K.P., 2007. The distribution of water ice in the interior
of comet Tempel 1.  Icarus 190, 284--294.

\bibitem[Toth et al.(2003)]{toth03} Toth, I., Lamy, P. L., Weaver,
H. A., 2003.  Hubble Space Telescope observations of the nucleus
fragment 73P/Schwassmann-Wachmann 3-B.  Bull. Amer. Astron. Soc.
35, 985. 

\bibitem[Toth et al.(2005)]{toth05} Toth, I., Lamy, P., Weaver, H.
A., 2005.  Hubble Space Telescope observations of the nucleus
fragment 73P/Schwassmann-Wachmann 3-C.  Icarus 178, 235--247.

\bibitem[Toth et al.(2006)]{toth06} Toth, I., Lamy, P., Weaver, H.,
A'Hearn, M., Kaasalainen, M., Lowry, S., 2006. Bull. Amer. Astron.
Soc.  38, 489. 

\bibitem[Villanueva et al.(2006)]{vill06} Villanueva, G. L., Bonev,
B.P., Mumma, M.J., Magee-Sauer, K., DiSanti, M. A., Salyk, C.,
Blake, G. A., 2006. The volatile composition of the split ecliptic
comet 73P/Schwassmann-Wachmann 3: A comparison of fragments C and
B.  Astrophys. J. 650, L87--L90.

\bibitem[Weaver et al.(2006)]{weaver06} Weaver, H. A., Lisse, C. M.,
Mutchler, M. J., Lamy, P., Toth, I., Reach, W. T., 2006.
Hubble Space Telescope investigation of the
disintegration of 73P/Schwassmann-Wachmann 3.
Bull. Amer. Astron. Soc. 38, 490. 

\bibitem[Weissman et al.(2006)]{weissman06} Weissman, P. R., Chesley,
S. R., Choi, Y. J., Bauer, J. M., Tegler, S.C., Romanishin, W.J.,
Consolmagno, G., Stansberry, J. A., 2006. Motion of the activity
source associated with active Centaur 174P/Echeclus (60558).  Bull.
Amer. Astron. Soc. 38, 551. 

\end{thebibliography}
\end{document}